\documentclass{article}

\usepackage{spconf,amsmath,graphicx, float, graphicx, booktabs, hyperref, mathrsfs, comment}
\usepackage{amsmath, amssymb}
\usepackage{multirow}
\usepackage{booktabs}
\usepackage{tabularx}
\usepackage{makecell}
\usepackage[linesnumbered,ruled,vlined]{algorithm2e}
\usepackage{enumitem}
\usepackage{url}
\usepackage{xcolor}
\usepackage{arydshln}
\setlist{nosep, leftmargin=14pt}

\usepackage[nolist]{acronym}
\usepackage{float}
\usepackage{mwe} 

\title{CASPER: Cross-modal Alignment of Spatial and single-cell Profiles for Expression Recovery}

\name{Amit Kumar $^{ \dagger}$ 
\qquad Maninder Kaur$^{\dagger}$ \qquad Raghvendra Mall $^{\star}$  \qquad  Sukrit Gupta$^{\ddagger}$}

\address{$^{\dagger}$  Department of Computer Science \& Engineering, Indian Institute of Technology Ropar, India\\
$^{\star}$ Qatar Computing Research Institute, Hamad Bin Khalifa University, Doha, Qatar.\\
$^{\ddagger}$ Department of Biomedical Engineering, Indian Institute of Technology Ropar, India }

\begin{document}
\ninept
\maketitle

\begin{acronym}[nolist]

    \acro{scRNA-seq}{Single-Cell RNA Sequencing}
    \acro{ST}{Spatial Transcriptomics}
    \acro{HVG}{Highly Variable Gene}
    \acro{MSE}{Mean Squared Error}
    \acro{MAE}{Mean Absolute Error}
    \acro{SRCC}{Spearman Rank Correlation Coefficient}
    \acro{MAPE}{Mean Absolute Percentage Error}
    \acro{CASPER}{Cross-modal Alignment of Spatial and single-cell Profiles for Expression Recovery}
    \acro{PCC}{Pearson Correlation Coefficient}
    \acro{LR}{Learning Rate}
    \acro{PCA}{Principal Component Analysis}
    \acro{VISc}{Visual cortex}
    \acro{SMSc}{Somatosensory cortex}
    \acro{POR}{Pre-optic region}
    \acro{SOTA}{state-of-the-art}
\end{acronym}

\begin{abstract}

\ac{ST} enables mapping of gene expression within its native tissue context, but current platforms measure only a limited set of genes due to experimental constraints and excessive costs. To overcome this, computational models integrate \ac{scRNA-seq} data with \ac{ST} to predict unmeasured genes. We propose \ac{CASPER}, a cross-attention based framework that predicts unmeasured gene expression in \ac{ST} by leveraging centroid-level representations from \ac{scRNA-seq}. We performed rigorous testing over four \ac{SOTA} \ac{ST}/\ac{scRNA-seq} dataset pairs across four existing baseline models. \ac{CASPER} shows significant improvement in nine out of the twelve metrics for our experiments. This work paves the way for further work in \ac{ST} to \ac{scRNA-seq} modality translation. The code for \ac{CASPER} is available at  \url{https://github.com/AI4Med-Lab/CASPER}.
\end{abstract}
\section{Introduction}
\label{sec:intro}

\ac{ST} has opened new possibilities for understanding how gene expression is organized within intact tissues, providing a bridge between molecular biology and histopathology. However, current \ac{ST} technologies, such as 10x Genomics' Visium and Xenium can measure only a few hundred selected genes due to assay and cost limitations, while most genes in the genome remain unmeasured. In contrast, \ac{scRNA-seq} captures genome-wide gene expression across individual cells but loses spatial context because tissues must be dissociated. Integrating these two complementary modalities has therefore become a key computational challenge in spatial omics research.


Recent studies have proposed several ways to bridge the gap between \ac{ST} and \ac{scRNA-seq} data. Probabilistic models such as gimVI \cite{gimVI} use variational inference to jointly model missing gene expression, while optimal transport methods such as SpaOTsc \cite{SpaOTsc} compute a global transport plan to match cells to spatial locations. Other approaches, including SpaGE \cite{SpaGE}, Tangram \cite{tangram2021deep}, and stPlus \cite{STPlus}, rely on \emph{global alignment}, where a single mapping or shared latent space is applied uniformly across the entire tissue to align the two modalities. Although effective in homogeneous settings, these global alignment strategies often over smooth the underlying biological variation and fail to capture region-specific transcriptional programs, especially in heterogeneous tissues. Methods such as STDiff \cite{STDiff} incorporate nonlinearity but still depend on global embeddings.

\begin{figure}[!ht]
\includegraphics[width=\columnwidth]{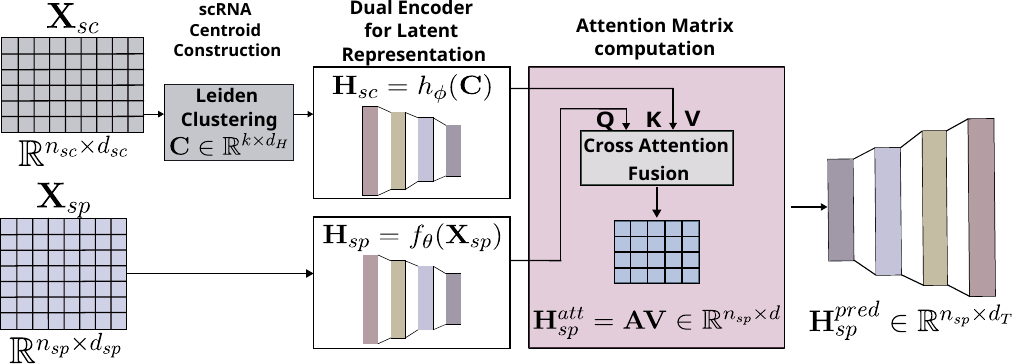}
\caption{
\textbf{Overview of the proposed CASPER framework for spatial gene expression imputation.}}
\label{fig:res}
\end{figure}

In this work, we introduce a new model, \ac{CASPER}, which imputes genome-wide gene expression for \ac{ST} using \ac{scRNA-seq} data. Briefly, the steps involved include, clustering \ac{scRNA-seq} cells into a small set of representative centroids using Leiden clustering \cite{traag2019louvain}, capturing distinct celltype transcriptional patterns while improving computational scalability. Then, a pair of modality-specific encoders projects spatial and \ac{scRNA-seq} data into a shared latent space, and a cross-attention decoder learns contextual correspondences by allowing each spatial spot to attend to the most relevant celltype centroids. We train an end-to-end model using masked spatial reconstruction as shown in Figure \ref{fig:res}. This design allows the model to transfer rich transcriptional information from \ac{scRNA-seq} to spatial locations while maintaining local consistency across tissue spots. Thus, \ac{CASPER} integrates ST and scRNA-seq data locally and contextually overcoming limitations of \ac{SOTA}.

Our primary contributions are summarized as follows:
\begin{itemize}
\item A method to encode each spatial gene expression vector to a distinct query vector that reflects its local transcriptional signature rather than applying a single global mapping across tissue.


\item Attention weights facilitate a straightforward biological interpretation 
of how each spatial location maps to a particular cell niche/cluster enabling efficient gene expression imputation.
\item Our method consistently outperforms \ac{SOTA} techniques in comprehensive comparisons across four different spatial transcriptomics datasets, providing significantly higher correlations (up to $0.230$) and significantly smaller errors, in some cases lowering MAPE by more than $60\%$.
\item Provide a foundation for spatial gene expression inference leveraging scRNA-seq across tissues and disease state.
\end{itemize}

\section{Methods}
Our goal is to predict genome-wide gene expression at spatial locations in the tissue, given that \ac{ST} data only has a subset of measured genes. For this, we design a cross-domain learning framework that leverages \ac{scRNA-seq} data as a high-dimensional reference. The \ac{scRNA-seq} dataset covers a broader transcriptional landscape than the \ac{ST} measurements.

\subsection{Problem Setup}
Let the spatial data be denoted by 
\(\mathbf{X}_{sp} \in \mathbb{R}^{n_{sp} \times d_{sp}}\), 
where \(n_{sp}\) is the number of spatial spots and \(d_{sp}\) is the number of genes profiled in the \ac{ST} platform.  
Similarly, the \ac{scRNA-seq} data is represented as 
\(\mathbf{X}_{sc} \in \mathbb{R}^{n_{sc} \times d_{sc}}\), 
with \(n_{sc}\) cells and \(d_{sc}\) measured genes.  
Only a subset of genes, denoted by \(\mathscr{G}_{shared}\), is common between the two modalities, and the target is the set of unmeasured genes \(\mathscr{G}_{target}\).  
Both matrices are log$_{1p}$-normalized and gene-wise standardized.

\subsection{scRNA Centroid Construction}
To reduce redundancy and computational cost while preserving biological diversity, we cluster scRNA cells using Leiden clustering \cite{traag2019louvain} on a \ac{PCA}-reduced graph constructed from \ac{HVG}.  
Each cluster \(c_i\) yields a centroid computed as the mean expression of its member cells:
\[
\mathbf{c}_i = \frac{1}{|C_i|}\sum_{\mathbf{x}_j \in C_i}\mathbf{x}_j,
\]
resulting in the centroid matrix \(\mathbf{C} = [\,\mathbf{c}_1^{\top},\, \mathbf{c}_2^{\top},\, \ldots,\, \mathbf{c}_k^{\top}\,]^{\top} \in \mathbb{R}^{k \times d_H}\) where \(k\) is the number of clusters and \(d_H\) is the \ac{HVG} dimension. Each centroid captures a celltype–level expression profile, acting as a compact biological token in the single-cell modality.


\subsection{Dual Encoders and Latent Representations}
Both modalities are embedded into a shared latent space of dimension \(d\) using separate feed-forward encoders:
\[
\mathbf{H}_{sp} = f_{\theta}(\mathbf{X}_{sp}), 
\quad 
\mathbf{H}_{sc} = h_{\phi}(\mathbf{C}),
\]
where 
\(\mathbf{H}_{sp} \in \mathbb{R}^{n_{sp} \times d}\)
and 
\(\mathbf{H}_{sc} \in \mathbb{R}^{k \times d}\).  
Each encoder consists of two fully connected layers with GELU activation and dropout:
\[
f_{\theta}(\mathbf{x}) = \mathrm{Linear}_2\!\big(\mathrm{Dropout}(\mathrm{GELU}(\mathrm{Linear}_1(\mathbf{x})))\big),
\]
and analogously for \(h_{\phi}\). These encoders learn modality-specific nonlinear transformations that preserve within-domain structure while projecting both domains into a comparable latent space.

\subsection{Cross-Attention Fusion}
We integrate \ac{ST} and \ac{scRNA-seq} representations using a cross-attention mechanism, 
which enables each spatial spot to selectively attend to relevant \ac{scRNA-seq} centroids.

\subsubsection{Query–Key–Value projections}
To form attention components, we learn linear projections:
\[
\mathbf{Q} = \mathbf{H}_{sp}\mathbf{W}_Q, \quad
\mathbf{K} = \mathbf{H}_{sc}\mathbf{W}_K, \quad
\mathbf{V} = \mathbf{H}_{sc}\mathbf{W}_V,
\]
where \(\mathbf{W}_Q, \mathbf{W}_K, \mathbf{W}_V \in \mathbb{R}^{d \times d}\) are learnable matrices, 
and thus 
\(\mathbf{Q} \in \mathbb{R}^{n_{sp} \times d}\), 
\(\mathbf{K}, \mathbf{V} \in \mathbb{R}^{k \times d}\).

\subsubsection{Attention computation}
Scaled dot-product attention between spatial queries and centroid keys is computed as:
\[
\mathbf{A} = \mathrm{softmax}\!\left(\frac{\mathbf{Q}\mathbf{K}^{\top}}{\sqrt{d}}\right),
\quad
\mathbf{A} \in \mathbb{R}^{n_{sp} \times k}.
\]
Each row of \(\mathbf{A}\) represents how a spatial spot distributes its attention over all \(k\) centroids.  
The attention-weighted fusion of centroid values yields the integrated spatial embeddings:
\[
\mathbf{H}_{sp}^{att} = \mathbf{A}\mathbf{V} \in \mathbb{R}^{n_{sp} \times d}.
\]

A residual connection and layer normalization are applied within the attention block to stabilize training. 
Finally, a linear projection maps the attention output to the target gene space:
\[
\mathbf{H}^{pred}_{\mathrm{sp}} 
= \mathbf{H}^{att}_{sp} W_{out} + \mathbf{b}_{out},
\quad
\mathbf{H}^{pred}_{sp} \in \mathbb{R}^{n_{sp} \times d_T},
\]
where $W_{out} \in \mathbb{R}^{d \times d_T}$ and $\mathbf{b}_{out} \in \mathbb{R}^{d_T}$ 
are learnable parameters, and $d_T = |\mathcal{G}_{\mathrm{target}}|$ denotes 
total target genes to be imputed.


\subsection{Loss Functions/ Masked ST reconstruction loss}
Only observed genes in the ST contribute to reconstruction:
\[
\mathscr{L}_{sp} 
= 
\frac{1}{\sum_{i,j} M_{ij}}
\big\|\mathbf{M}_{sp} \odot
(\mathbf{H}_{sp}^{pred} - \mathbf{H}_{sp}^{obs})\big\|_F^2,
\]

Here, $\mathbf{H}_{sp}^{pred} \in \mathbb{R}^{n_{sp} \times d_T}$ denotes the model-predicted expression matrix for all target genes, 
while $\mathbf{H}_{sp}^{obs}$ represents the subset of gene expressions that are experimentally observed in the \ac{ST} data. Since only a limited number of genes are measured in current \ac{ST} platforms, we apply a binary mask $\mathbf{M}_{sp} \in \{0,1\}^{n_{sp} \times d_T}$ to restrict the reconstruction loss to the observed entries. This masked mean-squared reconstruction term ensures that the model learns from valid spatial measurements while still being able to infer unmeasured genes during inference.


\subsection{Interpreting the results}
To show that \ac{CASPER} encapsulates, the one-to-many mapping of a spatial spot against the scRNA-seq based centroids, we look at the attention vector associated with each spot. If \ac{CASPER} mapped each spatial spot with multiple celltypes (forming clusters) from the input scRNA data, the attention scores would be diffused across the test samples. To measure this, we computed entropy and Top-2 gap of attention scores learnt by CASPER.


\noindent\textbf{Entropy}: From the attention matrix \(\mathbf{A}\), the attention vector for each spatial gene expression is given by \(\mathbf{a} = (a_1, a_2, \dots, a_K)\). The entropy of this probability distribution is given by:
\[E(\mathbf{a}) = -\sum_{i=1}^{K} a_i \log(a_i)\]
A large \(E(\mathbf{a})\) value indicates that the attention was spread across clusters. The maximum value that \(E(\mathbf{a})\) can take is upper-bounded by $\log(K)$, where $K$ is number of clusters.

\noindent\textbf{Top-2 Gap}: Let \(a_{(1)} \ge a_{(2)} \ge \cdots \ge a_{(K)}\) be the probabilities sorted in descending order. The Top-2 probability gap is given by:
\[F(\mathbf{a}) = a_{(1)} - a_{(2)}\]
A large gap indicates a distribution where one class clearly dominates, whereas a small gap indicates that the probability mass spread across at least two clusters. 



\section{Results}
\subsection{Datasets}
We evaluate the proposed framework using four publicly available \ac{ST} datasets (MERFISH/Moffitt \cite{merfish_d4}, seqFISH/AllenVISp \cite{seqFISH2019transcriptome_d2, allenVISp2018shared_d2}, osmFISH/Zeisel \cite{osmFish2018spatial_d1, zeisel2015cell_d1}, and STARmap/AllenVISp \cite{star2018three_d3, allenVISp2018shared_d2}), each paired with a corresponding \ac{scRNA-seq} reference dataset from the same or closely related mouse brain tissues. Each spatial dataset provides measured expression profiles for a subset of genes, typically ranging from tens to a few thousand, across spatially resolved tissue locations. These datasets encompass multiple spatial profiling technologies (osmFISH and MERFISH employ targeted hybridization, while seqFISH and STARmap achieve high multiplexing through sequential imaging), enabling us to assess the model's generalization across distinct platforms and spatial resolutions. 
The paired \ac{scRNA-seq} references contain transcriptomes of thousands of single cells with a broader gene coverage, serving as transcriptional priors for gene imputation. To ensure consistent preprocessing, all expression matrices were log$_{1p}$-normalized, and shared gene identifiers were harmonized between the \ac{ST} and \ac{scRNA-seq} modalities before downstream integration. 
The diversity in gene coverage and tissue structures across these datasets provides a rigorous benchmark for evaluating model robustness and cross-modal generalization. A detailed summary is provided in Table~\ref{tab:datasets}.

\begin{table}[!ht]
\scriptsize
\centering
\caption{Summary of \ac{ST} and \ac{scRNA-seq} datasets used. The datasets span multiple mouse brain regions, including the Pre-Optic Region (POR), Visual Cortex (VISc), and Somatosensory Cortex (SMSc).}
\label{tab:datasets}
\resizebox{.49\textwidth}{!}{
\begin{tabular}{c|ccc|cc}
\toprule
\textbf{Spatial/scRNA Data }& \textbf{Tissue} & \textbf{\# Spots} & \textbf{\#ST Genes} & \textbf{\# Cells} & \textbf{\# SC Genes} \\
\toprule
merFISH/Moffitt & POR & 64,373 & 155 & 31,299 & 18,646 \\
seqFISH/AllenVISp & VISc & 913 & 10,000 & 14,249 & 34,617 \\
osmFISH/Zeisel & SMSc & 3,405 & 33 & 1,691 & 15,075 \\
STARmap/AllenVISp & VISc & 1,549 & 1,020 & 14,249 & 34,617 \\
\bottomrule
\end{tabular}
}
\end{table}

\subsection{Dataset Split}


Each dataset was first divided into an 80\% training and a 20\% test split, where the test regions were kept spatially separate to prevent data leakage. Within the 80\% training portion, we further applied a 5-fold cross-validation scheme. For each fold, the shared gene set was partitioned such that a subset of genes were treated as targets and the remaining genes were used as model inputs during training. 

The model was trained on these training folds using matched \ac{scRNA-seq} centroids derived through \ac{HVG} selection, \ac{PCA}, and Leiden clustering, while the validation fold was used for early stopping and hyperparameter selection. During testing, we evaluated performance on the held-out 20\% spatial regions by predicting the same target genes that were removed during training, and we report results using \ac{PCC}, \ac{SRCC}, and \ac{MAPE} computed on these withheld genes.

\subsection{Model Training}


The model was trained end-to-end using the \texttt{Adam} optimizer with an initial learning rate of $10^{-4}$, dropout rate of 0.1, latent dimension \(d = 256\),  batch size of 128, and weight decay of $10^{-5}$ for a maximum of 200 epochs. $\mathscr{L}_{sp}$ is used as a loss function.

For efficiency, the scRNA embeddings \(\mathbf{H}_{sc}\) are precomputed at the beginning of each epoch.  
During training, each spatial batch attends to all \(k\) scRNA centroids via the attention matrix \(\mathbf{A}\), ensuring that every spatial region can access biologically relevant celltype information.  
At inference, the learned decoder maps the fused embeddings to \({\mathbf{H}}_{sp}^{pred}\), 
yielding log$_{1p}$-normalized predictions of unmeasured gene expression across spatial coordinates.


All experiments were implemented in the \texttt{PyTorch} framework using \texttt{Python 3.11} and executed on an NVIDIA RTX A5000 GPU (24 GB VRAM).

\subsection{Imputation performance comparison}

\begin{table*}[!h]
\small
\centering
\caption{Comparison of different imputation methods on the various datasets present in Table \ref{tab:datasets}, reported $\text{mean}_{\text{standard deviation}}$ across test sets. The \(*\) denotes statistically significant improvement (\(p < 0.05\), one-sided paired t-test) of the best-performing method (denoted by bold text) compared to the second best (underlined).}
\label{tab:perf_results}
\resizebox{\textwidth}{!}{
\begin{tabular}{lclccccc}
\toprule
\textbf{Dataset} &  \textbf{\shortstack{\# Common \\ Genes}} 
&\textbf{Metric} & SpaGE & Tangram & stPlus & stDiff & \textbf{CASPER} \\
\midrule

\multirow{4}{*}{\textbf{\shortstack{MERFISH/\\Moffitt}}} 
 &  \multirow{4}{*}{\centering 154}
 & Pearson $\uparrow$ & \underline{$0.334_{0.030}$} & $0.307_{0.039}$ & $0.302_{0.031}$ & $0.101_{0.017}$ & $\mathbf{0.503_{0.024}}^{*}$ \\
 &  & Spearman $\uparrow$ & $0.285_{0.034}$ & \underline{$0.286_{0.028}$} & $0.268_{0.033}$ & $0.097_{0.019}$ & $\mathbf{0.448_{0.022}}^{*}$ \\
  &  & MAPE (\%) $\downarrow$ & $147.140_{23.530}$ & \underline{$133.160_{22.950}$} & $142.900_{30.240}$ & $530.970_{207.300}$ & $\mathbf{72.790}_{9.940}^{*}$ \\
\midrule

\multirow{4}{*}{\textbf{\shortstack{seqFISH/\\AllenVISp}}} 
 &  \multirow{4}{*}{\centering 9,782}
 & Pearson $\uparrow$ & $0.096_{0.010}$ & $\mathbf{0.216_{0.002}}^{*}$ & $0.106_{0.005}$ & $0.001_{0.001}$ & \underline{$0.196_{0.003}$} \\
 &  & Spearman $\uparrow$ & $0.130_{0.010}$ & $\mathbf{0.303_{0.001}}^{*}$ & $0.128_{0.007}$ & $0.001_{0.001}$ & $\underline{0.301_{0.002}}$ \\
 &  & MAPE (\%) $\downarrow$ & $\underline{73.910_{0.240}}$ & $375.080_{301.320}$ & $274.680_{284.060}$ & $256.450_{4.410}$ & $\mathbf{65.950}_{0.470}^{*}$ \\
\midrule

\multirow{4}{*}{\textbf{\shortstack{osmFISH/\\Zeisel}}}
 &  \multirow{4}{*}{\centering 33}
 & Pearson $\uparrow$ & $0.150_{0.028}$ & $0.197_{0.033}$ & \underline{$0.199_{0.032}$} & $0.002_{0.011}$ & $\mathbf{0.429_{0.042}}^{*}$ \\
 &  & Spearman $\uparrow$ & $0.161_{0.033}$ & $0.165_{0.031}$ & \underline{$0.203_{0.042}$} & $-0.001_{0.016}$ & $\mathbf{0.406_{0.038}}^{*}$ \\
  &  & MAPE (\%) $\downarrow$ & \underline{$79.300_{4.300}$} & $84.160_{6.610}$ & $82.190_{6.910}$ & $130.880_{1.150}$ & $\mathbf{26.030}_{2.710}^{*}$ \\

\midrule

\multirow{4}{*}{\textbf{\shortstack{STARmap/\\AllenVISp}}}
 &  \multirow{4}{*}{\centering 994}
 & Pearson $\uparrow$ & $0.144_{0.013}$ & \underline{$0.176_{0.007}$} & $0.092_{0.011}$ & $0.025_{0.005}$ & $\mathbf{0.193_{0.007}}^{*}$ \\
 &  & Spearman $\uparrow$ & $0.135_{0.012}$ & \underline{$0.177_{0.007}$} & $0.100_{0.013}$ & $0.024_{0.004}$ & $\mathbf{0.197_{0.006}}^{*}$ \\
  &  & MAPE (\%) $\downarrow$ & $\mathbf{67.250}_{1.240}^{*}$ & $198.270_{131.190}$ & $158.040_{121.290}$ & $154.080_{4.880}$ & \underline{$74.420_{1.454}$} \\

\bottomrule
\end{tabular}
}
\end{table*}

Table \ref{tab:perf_results} presents a comprehensive comparison of the proposed method with four existing benchmarks including SpaGE, Tangram, stPlus, and stDiff across the selected datasets. From Table \ref{tab:perf_results}, we observed that \ac{CASPER} outperforms existing \ac{SOTA} across a majority of the metrics on myriad datasets. Besides the pearson and spearman correlation, we employed \ac{MAPE} to estimate the error in prediction by different models when compared to the groundtruth. \ac{CASPER} consistently (and significantly) gave lower \ac{MAPE} values (below 100 in all considered scenarios) as shown in Table \ref{tab:perf_results}.



\subsection{Interpretability results from CASPER}





\begin{figure}
    \centering
    \includegraphics[width=0.95\linewidth]{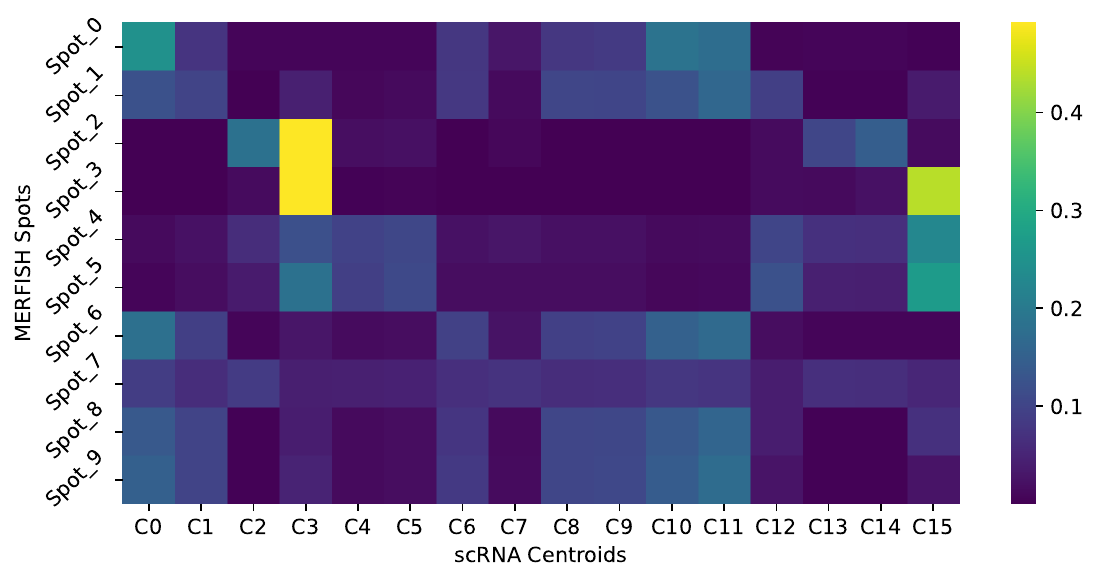}
    \caption{Attention-based mapping between merFISH ST spots ($\text{Spot}_x$) and scRNA-seq centroids (C$x$), where $x$ is variable.}
    \label{fig:attMap}
\end{figure}

The attention weights for a random set of merFISH \ac{ST} spots from the test set and the contributions from their corresponding \ac{scRNA-seq} centroids (obtained via Leiden clustering) are shown in Figure \ref{fig:attMap}. Figure \ref{fig:attMap} illustrates the degree to which a celltype contributes to the model's reconstruction of gene expression vector for a particular spatial location. Furthermore, Table \ref{table: interpretability} highlights the entropy and the Top-2 gap for each test set, where we consistently observe low entropy (minimum entropy is 0, maximum entropy is \(\log k\)) and high Top-2 gap indicating each spatial spot imputes gene expression predominantly from one scRNA-seq cluster centroid.

\begin{table}[!h]
\centering
\small
\resizebox{.5\textwidth}{!}{
\begin{tabular}{lclcc}
\hline
\textbf{Dataset} &\textbf{\# Centroids \((K)\)}  & \textbf{\(\log K\)} & \textbf{Entropy} & \textbf{Top-2 Gap} \\
\hline
merFISH  & 16  & 2.772 & $1.095_{0.555}$ & $0.588_{0.230}$\\ 
osmFISH  & 15  & 2.708 & $1.816_{0.273}$ & $0.090_{0.108}$\\ 
starMAP  & 29  & 3.367 & $0.485_{0.148}$ & $0.774_{0.073}$\\ 
\hline
\end{tabular}
}
\caption{Entropy and Top-2 Gap reported as  $\text{mean}_{\text{standard deviation}}$ across test sets from different datasets. We skip the results for the seqFISH dataset since the number of test samples are $<200$ with high sparsity for gene expressions.}
\label{table: interpretability}
\end{table}





\section{Conclusion}
In this work, we propose CASPER, a tranformer-based method that applies cross-attention between embeddings from \ac{ST} and \ac{scRNA-seq} data to impute missing gene expressions for \ac{ST}. 
Across all evaluated datasets, CASPER consistently outperforms existing baselines. Our analysis of the learned attention patterns confirms that CASPER does not rely on one-to-one cell matching; instead, it effectively weights information from multiple centroids (predominantly driven by a single cluster centroid) to reconstruct the missing gene expression for each spatial spot.


Overall, the results suggest that cross-attention provides a principled and inexpensive framework to couple \ac{ST} and \ac{scRNA-seq} data for imputing ST gene expression vector, paving the way for better downstream analysis and biomarker discovery with \ac{ST} data.



\bibliographystyle{IEEEbib}

\bibliography{strings,refs}

\end{document}